\documentstyle[seceq,preprint,epsf]{jpsj}

\def\runtitle{Enhancement of the Thermal Conductivity
in gapped Quantum Spin Chains}
\def\runauthor{Keiji {\sc Saito} and Seiji {\sc Miyashita}}

\title
{Enhancement of the Thermal Conductivity
 \protect \\  in gapped Quantum Spin Chains}

\author
{ Keiji {\sc Saito} and Seiji {\sc Miyashita} }

\inst
{Department of Applied Physics, Graduate School of Engineering
University of Tokyo, Bunkyo-ku, Tokyo 113-8656
}

\recdate
{
\today
}

\abst
{ We study mechanism of magnetic 
energy transport, motivated by recent measurements of the thermal conductivity 
in low dimensional quantum magnets. We point out 
a possible mechanism of enhancement of the thermal conductivity 
in gapped magnetic system, where the magnetic energy transport plays a crucial role.
This mechanism gives an interpretation for 
the recent experiment of CuGeO$_{3}$, 
where the thermal conductivity depends on the crystal direction.
}

\kword
{Quantum Magnetism, Thermal Conductivity}

\begin{document}
\sloppy
\maketitle
\section{Introduction}
The relationship of the thermal conductivity and magnetic state 
in various low dimensional quantum magnets has been attracted interests
\cite{SFGOVR00,KNKNK01,HLUKZFKU01,KINAKMTK01,SGOARBT00,SGOAR00}.
The thermal conductivity can be a strong experimental instrument to
investigate the magnetic properties in materials. 
For the magnetic systems with large exchange coupling constants, the 
magnetic energy transport dominantly contributes to the thermal 
conductivity. Low-dimensional spin systems with 
many conserved quantities tend to possess a ballistic energy transport,
which means an infinite thermal conductivity 
as studied in the theoretical side \cite{STM96,Z99,FK98}. 
These facts can give a possible mechanism of a 
large magnetic thermal conductivity.
Solgubenko et al. observed such nondiffusive magnetic 
energy transport in SrCuO$_2$ and Sr$_{2}$CuO$_{3}$ 
which are described by the quasi-one dimensional 
isotropic antiferromagnetic Heisenberg model \cite{SFGOVR00}.
These materials have very large exchange coupling constant 
$J\sim 2000 K$ so that the magnon runs through the system even at 100 K. 
They found that the energy is transmitted by magnon at this temperature.

Unusual enhancement of conductivity is observed 
at low temperatures below a spin gap temperature 
in gapped spin system such as the two-dimensional dimer spin
system SrCu$_{2}$(BO$_{3}$)$_{2}$, and the ladder system 
(Sr, Ca)$_{14}$Cu$_{24}$O$_{41}$
\cite{KNKNK01,HLUKZFKU01,KINAKMTK01,SGOARBT00,SGOAR00}. 
This enhancement is attributed to the contribution of phononic
energy transport. This phononic energy transport 
is quantitatively explained in view of `resonant scattering 
process' between phonon and spin gap \cite{HLUKZFKU01,SGOAR00,NV72}.
That is, the magnetic system is excited absorbing 
a phononic energy, and another phonon of the same energy is emitted 
deexciting the magnetic system at the same time \cite{NV72}.
These experiments suggest that
at high temperatures the magnetic energy transport dominantly convey the heat, 
while at low temperatures
the magnetic gap suppresses the magnetic energy transport.
Thus phononic energy transport become dominant. Therefore
the energy carrier changes from
spins to phonons around the spin gap temperature. 

The spin-Peierls (SP) system is another gapped system where the energy gap 
is formed below a critical temperature $T_{\rm SP}$. 
CuGeO$_{3}$ is a typical material which shows the SP transition. 
Ando et al. studied the thermal conductivity in the chain direction in 
CuGeO$_{3}$ and found the enhancement of the conductivity below 
$T_{\rm SP}$ \cite{ATSDTFK98} which is similar to that
found in the two-dimensional dimer and ladder systems 
\cite{KNKNK01,HLUKZFKU01,KINAKMTK01,SGOARBT00,SGOAR00}.
Ando et al. attributed this unusual enhancement of
conductivity to the increase of the phononic transport 
due to the decrease of spin-phonon coupling interaction.  
The similar enhancement is also observed more clearly 
in NaV$_{2}$O$_{5}$ where  the charge ordering transition 
occurs \cite{VPKDRIU98,PVDRIU99}.

Recently Hofmann et al. compared the properties of thermal conductivity of 
SrCu$_{2}$(BO$_{3}$)$_{2}$ and CuGeO$_{3}$ \cite{HLFUKUDR}. They investigated 
the thermal conductivity along different crystal axes,
although the previous experiment in CuGeO$_{3}$ was done only in the
chain direction \cite{ATSDTFK98}. 
Generally temperature dependence of phononic thermal 
conductivity does not depend on the crystal direction,
whereas a magnetic one does on it. Thus by studying the 
crystal direction dependence of thermal conductivity, 
whether the enhancement of the thermal conductivity 
is caused by phonons or not is clarified. 
SrCu$_{2}$(BO$_{3}$)$_{2}$ is described by the Shastry-Sutherland 
model \cite{SS81} which has the exact dimerized ground state and 
almost dispersionless triplet excitations where the magnetization is localized.
Thus below the spin gap temperature, we expect that the 
phononic energy transport is dominant because of small magnetic contribution.
Actually this prospect is supported by the observation that the thermal 
conductivity does not depend on the crystal direction.
On the other hand, they found that CuGeO$_{3}$ shows the direction dependence
of thermal conductivity. 
%This indicates that magnetic thermal conductivity 
%contributes to the enhancement against the previous belief
%\cite{ATSDTFK98}.

In this paper, motivated by these recent experiments, we theoretically 
investigate the role of magnetic transport when the system has the energy gap,
and propose a mechanism of enhancement of the thermal conductivity 
in magnetic systems. We consider a bond-alternating antiferomagnetc 
Heisenberg chain, and investigate
the thermal conductivity focusing on the dependence on the energy gap.
By the quantum master equation approach,
we study a spin system which is connected to two reservoirs of 
different temperatures.
We found that the 
thermal conductivity is enhanced due to the presence of bond-alternation
around the spin gap temperature.
This observation gives an interpretation that a small bond-alternation 
yields an enhancement of specific heat but does not drastically
change the mean free path. This interpretation 
is consistent with Hofmann's recent experiments \cite{HLFUKUDR}.

This paper is organized as follows. In \S $2$, we introduce the model, 
and the method to investigate the thermal conduction is
explained in \S $3$.
Summary and discussions are given in \S $4$.

\section{Model}
The system we shall consider is an alternate Heisenberg spin chain 
described by,
\begin{eqnarray}
{\cal H}_{s} &=& J \sum_{\ell =1}^{N -1} \left( 1 - (-1)^{\ell} \delta \right)
{\bf S}_{\ell} \cdot {\bf S}_{\ell +1} , \label{hamil}
\end{eqnarray}
where $J$ is an exchange interaction which is taken as a unity throughout 
this paper, and ${\bf S}_{\ell}$ is the $\ell$th spin.
The number of spins $N$ is taken to be even.
A finite value of parameter $\delta$ causes a bond-alternation, which
yields the dimerized ground state and 
a finite spin gap \cite{dimer}. 
We study the thermal conductivity for various values of 
$\delta$.
	
The thermal conductivity $\kappa ( T )$ at the temperature 
$T$ is generally given by the Green-Kubo formula which reads as,
\begin{eqnarray}
\kappa ( T ) &=& 
\lim_{t\to \infty}\lim_{N\to\infty}\frac{1}{2N T^2} \int_{0}^{t} \, du \, 
\langle\left\{ \hat{J}, \hat{J}(u)\right\}\rangle , \label{gk}
\end{eqnarray}
where $\left\{ ., .\right\}$ means the anti-commutation relation, and 
$\langle ... \rangle$ is the equilibrium average at the temperature $T$.
The operator $\hat{J}(u)$ is the total heat current operator at the time $u$ 
in the Heisenberg picture, i.e., 
$\hat{J}(u) = \exp(i {\cal H} u) \hat{J} \exp(-i {\cal H} u)$. 
The total current operator $\hat{J}$ for (\ref{hamil}) is calculated
by the continuity equation of the energy as,
\begin{eqnarray}
\hat{J} &=& -i J\sum_{\ell =1}^{N-2} \left( 1- \delta^2 \right)
\left[ {\bf S}_{\ell} \cdot {\bf S}_{\ell +1} , \, {\bf S}_{\ell +1} 
\cdot {\bf S}_{\ell +2} \right] \nonumber \\
&=& - J \left( 1- \delta^2 \right) 
\sum_{\ell =1}^{N-2}  \left( {\bf S}_{\ell } \times 
{\bf S}_{\ell +1}\right) \cdot {\bf S}_{\ell +2}  \label{tc}
\end{eqnarray}
where $\left[ ., . \right]$ means the commutation relation.
In the case of the isotropic Heisenberg model with the periodic boundary 
condition, i.e. $\delta =0$, the total current (\ref{tc}) 
is a conserved quantity.
Therefore the Green-Kubo formula (\ref{gk}) diverges, which 
means that magnons are not scattered and the mean free path of
each magnon diverges. 
In real magnetic materials, the relaxation time becomes finite 
due to the other sources such as spin-phonon
scattering and impurities, and so on.

In order to take these sources into account for the isotropic case 
$\delta =0$, more practical formulation of the thermal conductivity 
is used
especially when the magnetic low excitation contributes to the energy 
transport at low temperatures \cite{SFGOVR00,FT81}; 
\begin{eqnarray}
\kappa(T) &=& v_{\rm G}^2 \int \, dk \, {\cal D}(k) \tau (k) \frac{d}{dT} 
\left[\frac{\hbar\omega (k)}{e^{\beta \hbar\omega (k) } +1 } \right] 
\label{kp}, 
\end{eqnarray}
where $\omega (k)$ is the low energy dispersion
which is taken to be the 
Des Cloizeaux-Pearson mode \cite{DP62} 
$\omega (k) =\frac{J\pi}{2}\sin \,k$.
$v_{\rm G}$ is the group velocity.
${\cal D}(k)$ and $\tau (k)$ are the density 
and the relaxation time of the mode $k$, respectively.
Eq. (\ref{kp}) can be also written as 
$\kappa = \sum_{k} C_{k}(T) D_{k}(T)$ where $C_{k}(T) = {\cal D}(k) 
\frac{d}{dT}  
\left[\frac{\hbar\omega (k)}{e^{\beta \hbar\omega (k) } +1 } \right] 
$ and $D_{k}(T) = v_{\rm G}^2 \tau (k) $. $C_{k}(T)$ and $D_{k}(T)$
are interpreted as 
the specific heat and the diffusion constant of energy for the mode $k$,
respectively. 
This is derived with the same spirit in the phononic 
transport discussed by the Peierls-Boltzmann equation \cite{Ktext}.
The relaxation time $\tau (k)$ is determined by various 
origins such as impurities and spin-phonon coupling, and so on. 
Practically, the relaxation rate is calculated by the Carraway's 
method \cite{C59}, i.e., by the summation over relaxation 
rates originated from these origins.
In this formula, the mean free path $l_{k}$ is defined using the 
relaxation time $\tau (k)$ as $l_{k} = v_{\rm G} \tau (k)$. 
In CuGeO$_{3}$, above $T_{\rm SP}$ where the Hamiltonian 
is roughly regarded as the isotropic Heisenberg chain $\delta = 0$, 
the mean free path is estimated as $500$ lattice size \cite{ATSDTFK98}. 
We should note that 
the mean free path is a dynamical property and is different 
from the spin correlation length in the equilibrium state.

In order to apply the formula (\ref{kp}) to the case of finite $\delta$, 
we must know the low energy dispersion with finite energy gap, 
although we do not know exact analytical expression for it.
A finite energy gap causes the localization of the magnetic excitation 
which suppresses the energy diffusion, whereas energy gap enhances
the heat capacity around spin gap temperature.
For instance, when $\delta =1$, the system is completely 
separated into $N/2$ local spin pairs, and
the group velocity is $0$.
Therefore this extreme case gives zero thermal conductivity
due to vanishing mean free path.
Thus it is not trivial whether the existence of energy gap 
enhances thermal conductivity or not. In this context, we will
study the thermal conduction by a direct numerical method.

\section{Method and Results}
We investigate how the thermal conductivity behaves as a function of 
$\delta $.
It is difficult to treat the Green-Kubo formula (\ref{gk}) numerically 
because it requires infinite $N$. 
Therefore the quantum master equation approach is more tractable 
in numerical simulations.
We study the spin system which is directly connected to
the reservoirs of different 
temperatures at the ends.
We here consider a system of $N=6$ and $N=8$. Although this lattice sizes are small, 
we believe that the essential mechanism of magnetic energy transport 
is clarified.
The time-evolution of the system is determined by
the following master equation
 (e.g., see the reference \cite{STM00} and references therein);
\begin{equation}
\frac{\partial\rho(t)}{\partial t} = 
-i\left[ {\cal H}, \rho (t) \right] -\lambda 
\left( {\cal L}_{\rm L} \rho (t)+ {\cal L}_{\rm R} \rho (t) \right) ,
\label{QMEq}
\end{equation}
where the first term in the right-hand side corresponds to 
the pure quantum dynamics of the system, and 
${\cal L}_{\rm  L}$ and ${\cal L}_{\rm  R}$ express
the contact with the thermal reservoirs of the inverse temperature $\beta_{\rm L}
(=1/T_{\rm L})$ at the left end and $\beta_{\rm R} (=1/T_{\rm
R})$ at the right end, respectively. The parameter $\lambda$ is the coupling strength.
The dissipative term $ {\cal L}_{\alpha } \, (\alpha =
{\rm L},{\rm R})$ is given by
\begin{eqnarray}
 {\cal L}_{\alpha } \rho(t) =
\left( \left[X_{\alpha }, R_{\alpha } \rho(t)\right] 
+ \left[X_{\alpha }, R_{\alpha } \rho(t) \right]^{\dag} \right) ,
\end{eqnarray}
where $X_{\rm L}$ and $X_{\rm R}$ are the system's operators 
directly attached to the left and right reservoir, respectively.
Here we take $X_{\rm L} = S_{1}^{z}$ and $X_{\rm R } =
S_{N}^{z}$.
The operator $R_{\alpha }$ is given by 
$\langle k | R_{\alpha }  | m \rangle = 
(E_{k} - E_{m}) ({e^{\beta_{\alpha } (E_{k} - E_{m})} -1 })^{-1}
\langle k | X_{\alpha }  | m \rangle $ in the representation 
diagonalizing the Hamiltonian (\ref{hamil}) as ${\cal H} | k \rangle =
E_{k} | k \rangle $ and ${\cal H} | m \rangle = E_{m} | m \rangle $.
If $\beta_{\rm L} = \beta_{\rm R}$, 
the master equation (\ref{QMEq})
gives the canonical distribution as the stationary solution, which is
easily checked by substituting the canonical 
distribution for the density matrix $\rho(t)$. 
We numerically integrate the equation (\ref{QMEq}) 
and obtain the stationary density matrix $\rho_{\rm st}$ starting from
the initial density matrix of the canonical distribution at 
temperature $T_{\rm R}$.
The simulation was carried out 
by the fourth order Runge-Kutta method with the time step $0.01$ 
with $\lambda =0.01$. 

We define the thermal conductivity by measuring the stationary energy
current ${\rm Tr }\left( \rho_{\rm st} \hat{J}\right)$ setting 
\begin{eqnarray}
T_{\rm R} = T_{\rm L} + \Delta T ,
\end{eqnarray}
where $\Delta T$ is a temperature difference between two reservoirs, and
is taken as $\Delta T = 0.3$. Thereby the thermal 
conductivity $\kappa (T)$ is defined as
\begin{eqnarray}
\kappa ( T= T_{\rm L}) = 
{{\rm Tr} \left( \hat{J} \rho_{\rm st} \right) \over \Delta T}.
\end{eqnarray}

We present the results 
of the thermal current $\kappa (T) \Delta T $ normalized by $N-2$,
which is the number of terms in the summation (\ref{tc})
In Fig.1(a) and (b), the normalized thermal currents are shown for $N=6$
and $N=8$, respectively.
In the inset, the data of the specific heat are also shown. 
The normalized thermal current show little quantitative difference
between the cases of $N=6$ and $N=8$. 
In the figures,
we obtained finite thermal
conductivities even in the isotropic case, $\delta =0$. 
As explained in previous section, it must diverge in
the Green-Kubo formula. 
These finite values of thermal conductivity
are attributed to the finite size effect 
in the presence of the thermal reservoirs at ends of the system, 
and the conductivities must diverge in the limit 
of $N\rightarrow \infty$. 
Actually the conductivity for $N=8$
is larger than that for $N=6$. This size dependence is also observed for
the cases of finite values of $\delta$. 

In the isotropic case, $\delta =0$, 
the overall form of the thermal conductivity 
is similar to that obtained by Kl\"{u}mper and Sakai who 
exactly calculated the amplitude of 
zero frequency contribution in the Green-Kubo formula \cite{KS01}.
That is, the thermal conductivity has one peak at about $T \sim 0.5$.
This behavior is a common characteristic which does not change 
when the system size increases. 
At very low temperatures $\kappa(T)$ should be in proportional to $T$
due to the temperature dependence of specific heat, which is
similar to the Casimir's theory in phononic transport \cite{C38}.
This low temperature property, however, cannot be checked 
for such finite sizes.  
We must note that even in the isotropic case $\delta =0$, the system
has a finite energy gap due to the finite size effect.
When $\delta$ becomes finite, the gap increases, e.g., in the case of
$N=8$, $\Delta E = 0.39269, \Delta E =  0.74750,  1.09041,  1.41106,  
1.71290$, and $1.85807$ for $\delta = 0.0,0.2,0.4,0.6,0.8$, and $0.9$,
respectively.

Small bond alternations enhance the thermal conductivity
in spite that the alternation tends to separate the system into local 
spin pairs. This enhancement is observed for even $\delta =0.8$, and 
this feature is quite robust. 
In the case of very large $\delta $, the thermal conductivity is reduced 
due to the effect of separation of the system. The temperature of the 
peak of thermal conduction roughly corresponds to half of energy gap like
the Shotky-type specific heat. Actually the overall behavior of
thermal conductivity is similar to that of the specific heat.
Thus from this observation, we conclude that 
the mean free path does not drastically changed 
by the bond-alternation. 
We expect that the unusual enhancement of the thermal conductivity in
the spin-Peierls system is caused by magnetic energy transport 
when $T_{\rm SP}$ is the same order of $\Delta E$.
That is, in the spin-Peierls system, 
$\kappa (T)$ follows the data of $\delta =0$ at the high temparture
region, whereas below $T_{\rm SP}$, $\delta$ suddenly increases and
$\kappa(T)$ is given by $\kappa( T , \delta\ne0 )$, which gives a sharp
peak when $T_{\rm SP}$ is the same order of $\Delta E$. 
In CuGeO$_{3}$, $T_{\rm SP}$ is the same order of $\Delta E$, i.e.,
$T_{\rm SP} \sim 14 K$ and $\Delta E \sim 20 K$.

\section{Discussion}
Magnetic energy transport is becoming a strong instrument to capture the
magnetic properties of quantum magnets.
In order to realize an accurate interpretation for obtained 
experimental data, theoretical arguments are necessary. 
We demonstrate how the thermal conductivity is enhanced due to the magnetic
interaction, and 
proposed that besides the enhancement due to the increase of 
phononic energy transport, magnetic energy transport also causes an
enhancement at a temperature of the order of $\Delta E$.
This enhancement is directly related to
that of the specific heat. This correspondence indicates 
that the mean free path does not drastically changed by a small
bond-alternation.
If the bond-alternation disturbs the dynamics 
of the system and the mean free path becomes much smaller than the isotropic 
chain, this correspondence between the thermal conductivity and specific heat
is no more valid. The numerical simulation, however,
indicates that there is some region where this mechanism of enhancement
really exists.
We hope that this effect by magnetic energy transport is also
studied in experiment.

So far, it is difficult to control the heat flow in material.
However making use of the magnetic energy transport, we 
may plan a new device that controls the energy flow. In such device,
the energy flow could be drastically changed by controlling some parameters
such as external magnetic fields. 
The creation of a energy gap provides a promising 
mechanism which enhances the energy flow. 
It is interesting to study 
mechanisms to control magnetic transport in more detail, which will be reported
elsewhere.

\section*{Acknowledgement}
The computer calculation 
was partially carried out at the computer
center of the ISSP, which is gratefully acknowledged.
The present work is supported by Grand-in-Aid for Scientific 
Research from Ministry of Education, Culture, Sports, Science,
and Technology of Japan.

\newpage
\begin{figure}
\begin{center}
\noindent
\epsfxsize=9cm \epsfysize=7cm
\epsfbox{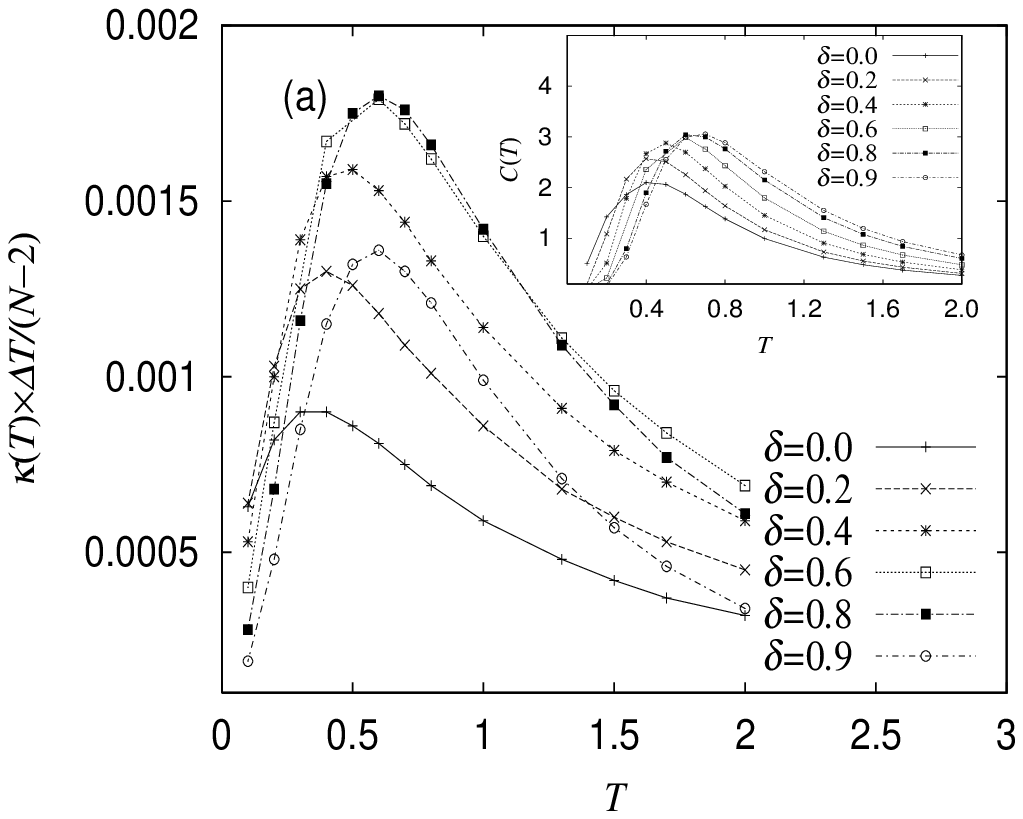} 
\epsfxsize=9cm \epsfysize=7cm
\epsfbox{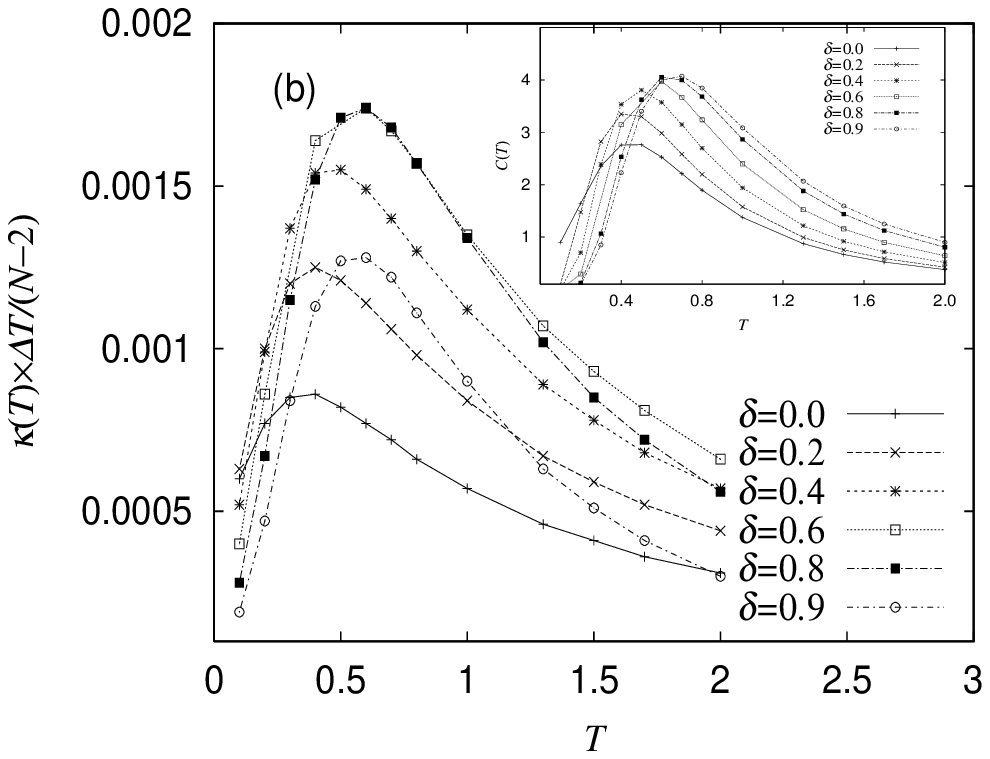}
\caption{The thermal conductivities for various $\delta$ with the system size,
(a): $N=6$ and  (b): $N=8$.
The inset is the specific heat. 
}
\end{center}
\end{figure}
\begin{center}
K. Saito and S. Miyashita
\end{center}

\end{document}